\documentclass{article}
\usepackage{ltwol2e}

\usepackage{epsfig}

\def\NPB{{\em Nucl. Phys.} B}
\def\PLB{{\em Phys. Lett.}  B}

\def\ZPC{{\em Z. Phys.} C}

\def\be{\begin{equation}}
\def\ee{\end{equation}}
\def\bea{\begin{eqnarray}}
\def\eea{\end{eqnarray}}
\newcommand{\roots}{\ensuremath{\sqrt{s}}}
\newcommand{\WW}{\mbox{$\mathrm{W^+W^-}$}}
\newcommand{\Mw}{\mbox{$M_{\mathrm{W}}$}}
\newcommand{\epem}{\ensuremath{\mathrm{e}^+\mathrm{e}^-}}
\newcommand{\DQQQQ}{\mbox{$\mathrm{D^{\mathrm{4q}}}$}}
\newcommand{\DQQLV}
{\mbox{$\mathrm{D}^{\mathrm{q\overline{q}}\ell\overline{\nu}_{\ell}}$}}
\newcommand{\DD}{\mbox{$\Delta\mathrm{D}$}}
\newcommand{\nQQQQ}{\mbox{$\langle n_{\mathrm{ch}}^{\mathrm{4q}}\rangle$}}
\newcommand{\nQQLV}{\mbox{$\langle n_{\mathrm{ch}}^{\mathrm{qq}\ell\nu}\rangle$}}
\newcommand{\Dnch}{\mbox{$\Delta\langle n_{\mathrm{ch}}\rangle$}}
\newcommand{\WWqqqq}{\mbox{\WW$\rightarrow$\qq\qq}}
\newcommand{\WWqqlv}{\mbox{\WW$\rightarrow$\qq\lnu}}
\newcommand{\qq}{\mbox{$\mathrm{q\overline{q}}$}}
\newcommand{\lnu}{\mbox{$\ell\overline{\nu}_{\ell}$}}
\newcommand{\Zz}{\ensuremath{{\mathrm{Z}^0}}}
\newcommand{\Gw}{\ensuremath{\Gamma_{\mathrm{W}}}}
\newcommand{\Zgamma}{\mbox{$\Zz/\gamma$}}
\newcommand{\Zqq}{\ensuremath{\Zz/\gamma\rightarrow\qq}}
\newcommand{\xpQQQQ}{\mbox{$\langle x_{p}^{\mathrm{4q}}\rangle$}}
\newcommand{\xpQQLV}
{\mbox{$\langle x_{p}^{\mathrm{q\overline{q}}\ell\overline{\nu}_{\ell}}\rangle$}}
\newcommand{\Dxp}{\mbox{$\Delta \langle x_{p} \rangle$}}
\newcommand{\thrQQQQ}{\mbox{$\langle (1-T)^{\mathrm{4q}}\rangle$}}
\newcommand{\yQQQQ}{\mbox{$\langle |y^{\mathrm{4q}}|\rangle$}}
\newcommand{\SK}{\mbox{Sj\"ostrand-Khoze}}
\newcommand{\SKI}{\mbox{SK I}}
\newcommand{\SKII}{\mbox{SK II}}
\newcommand{\SKIIpr}{\mbox{SK II$^{\prime}$}}
\newcommand{\ARMII}{\mbox{AR 2}}
\newcommand{\ARMIII}{\mbox{AR 3}}

\newcommand{\EG}{\mbox{Ellis-Geiger}}

\newcommand{\ARIADNE}{\mbox{A{\sc riadne}}}
\newcommand{\JETSET}{\mbox{J{\sc etset}}}
\newcommand{\VNI}{\mbox{V{\sc ni}}}
\newcommand{\HERWIG}{\mbox{H{\sc erwig}}}
\newcommand{\KORALW}{\mbox{K{\sc oralw}}}

\newcommand{\nPLB}[3]  {\PLB\ \textbf{#1} (#2) #3}
\newcommand{\nZPC}[3]  {\ZPC\ \textbf{#1} (#2) #3}
\newcommand{\nNPB}[3]  {\NPB\ \textbf{#1} (#2) #3}
\newcommand{\nPRD}[3]  {Phys.\ Rev.\ \textbf{D#1} (#2) #3}
\newcommand{\nCPC}[3]  {Comp.\ Phys.\ Comm.\ \textbf{#1} (#2) #3}

\def\etal{\mbox{{\it et al.}}}
\def\gappeq{\ensuremath{\mathrel{ \rlap{\raise.5ex\hbox{>}}
                      {\lower.5ex\hbox{\sim}}}}}
\def\lappeq{\ensuremath{\mathrel{ \rlap{\raise.5ex\hbox{<}}
                      {\lower.5ex\hbox{\sim}}}}}

\begin{document}
\title{\begin{flushright}
  BHAM-HEP/98-01 \\
  28th September 1998
\end{flushright}
\WW\ HADRONIC DECAY PROPERTIES}

\author{N. K. WATSON}

\address{School of Physics and Astronomy, University of Birmingham,
P.O. Box 363, Birmingham B15 2TT, United Kingdom\\
Invited talk at ICHEP'98, $22-30^{th}$ July 1998, Vancouver, Canada.}


\twocolumn[\maketitle\abstracts{
Recent measurements of the properties of \WW\ events
produced in \epem\ collisions at $\roots\sim183$~GeV at LEP
are reviewed. The data are used to investigate the predicted
effects of final state interactions, specifically ``colour
reconnection''. }]

\section{Motivation}
\noindent 
Hadronic data in \epem\ collisions can be characterised by event shape
distributions and inclusive observables such as charged particle
multiplicities and momentum spectra.  In addition to tests of Monte
Carlo models, measurement of the properties of the hadronic sector of
\WW\ decays allows the question of `colour reconnection'
\cite{bib:GPZ} to be addressed experimentally. The decay products of
the two W decays may have a significant space-time overlap as the
separation of their decay vertices at LEP2 energies is small compared
to characteristic hadronic distance scales. In the fully hadronic
channel this may lead to final state interactions. Colour reconnection
is the general name applied to the case where such final state interactions
lead to a rearrangement of the colour flow between the two W bosons.
At present there is general consensus that observable
effects of interactions between the colour singlets during the
perturbative phase are expected to be small.  In contrast, significant
interference in the hadronisation process is considered to be a real
possibility.  With the current knowledge of non-perturbative QCD, such
interference can be estimated only in the context of specific
models.~\cite{bib:GPZ,bib:SK+GH+HERWIG,bib:ARIADNE,bib:EG}
One should be aware that other final state effects
such as Bose-Einstein correlations between identical bosons
may also influence the observed event properties. The combined action of
these two effects may be either to reduce or enhance possible 
characteristic signatures of their presence.

\section{Event Properties}
The results shown are based on $\sim 55$~pb$^{-1}$ data per LEP
collaboration.~\cite{bib:lep_vanc_props}  Simple observables such as the
inclusive charged multiplicity are obvious candidates for study. There have
been predictions \cite{bib:EG} that the effects of colour reconnection may
be $\sim$ 10\% on \nQQQQ, the mean charged multiplicity in \WWqqqq\
events. The reference sample against which such changes are gauged is
typically taken to be twice the multiplicity of the hadronically decaying W
in \WWqqlv\ events, \nQQLV. Charged particles associated with the leptonic
component of such events are excluded.  The difference is defined
$\Dnch=\nQQQQ-2\nQQLV$. This reduces the dependence on the modelling of
single W decays at the expense of a lower statistical significance of
the test performed.

\begin{figure}
\center
\psfig{figure=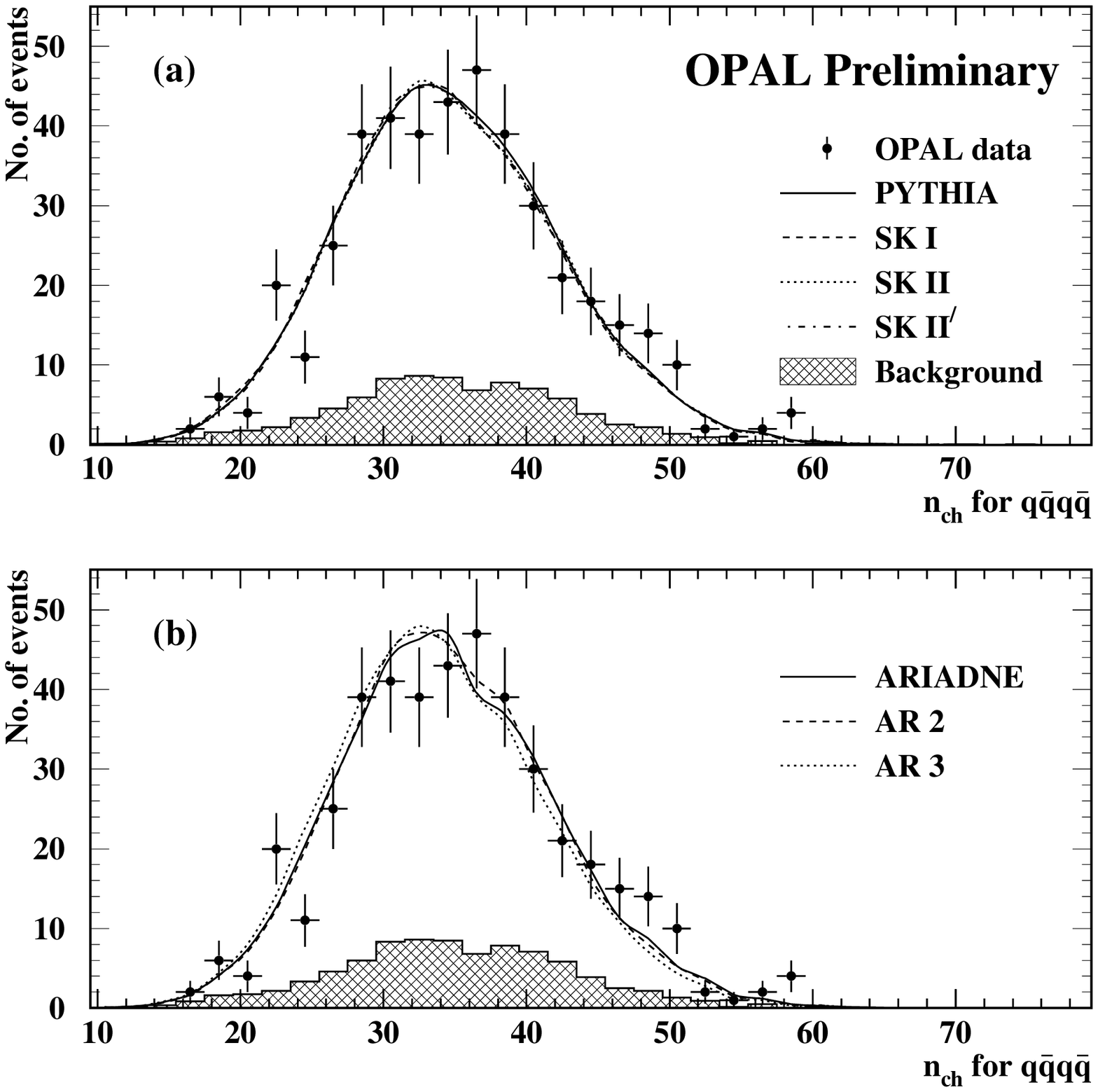,width=8cm}
\caption{Observed track multiplicity, \WWqqqq\ events.}
\label{fig:nch_opal}
\end{figure}

\noindent
The observed (uncorrected) track multiplicities for selected \WWqqqq\
candidates are shown in Figure~\ref{fig:nch_opal} by OPAL, together with
the predictions from a variety of colour reconnection models.  The mean
charged particle multiplicities, \nQQQQ\ and \nQQLV, may be extracted from
such data, after subtraction of the predicted background, by employing a
matrix-based unfolding procedure. This uses the event-by-event correlation,
taken from Monte Carlo, between the charged multiplicity at hadron level and
that observed in the detector after all analysis cuts have been applied. A
subsequent correction for the effects of initial state photon radiation and
to full acceptance is applied.  Alternative methods for measuring \nQQQQ\
and \nQQLV\ are the integration of any charged particle event shape, such as
the distributions of momentum, rapidity or transverse momentum.  Each
collaboration uses more than one method to control possible systematic
effects.

Changes in \nQQQQ\ predicted vary with the colour reconnection model used,
and the extent to which it may have been retuned.  Representative shifts,
defined as $\nQQQQ^\mathrm{reconnection}- \nQQQQ^\mathrm{normal}$ for each
model, are:

\begin{eqnarray*}
  \SK\ \mathrm{(SK)\; I}            &\sim& -0.3 \\
  \SKII, \SKIIpr\                   &\sim& -0.2 \\
  \mathrm{\ARIADNE\ 2   }           &\sim& -0.1 \\
  \mathrm{\ARIADNE\ 3 \; (\ARMIII)} &\sim& -1.0
\end{eqnarray*}

\noindent
As unfolding the data may introduce a bias towards the model used, the
procedure is repeated using a variety of models to estimate any associated
systematic effect.  L3 and OPAL also unfold their data using several colour
reconnection models as part of such studies.  ALEPH correct their data for
experimental effects, such as finite detector resolution, but not for losses
due to tracking inefficiency in the low momentum region ($p_T<200$~MeV), or
phase space and topological biases invariably present in experimental event
selections. By using only a partial unfolding, their data should be less
affected by model biases but cannot be compared directly with other
experimental results, or to model predictions without passing through their
detector simulation and analysis cuts.
\begin{table}
\begin{center}
\caption{Mean charged particle multiplicities, errors are
sum of statistical and systematic components.
$\dag$: only partially unfolded.}\label{tab:nch}
\vspace{0.2cm}
\begin{tabular}{|l|c|c|c|} \hline
  Expt. &  \nQQQQ  &  \nQQLV &  \Dnch \\
\hline
ALEPH$^\dag$
  &  $35.33\pm0.73$ &  $17.01\pm0.37$  &
                                  $+1.31\pm0.83$ \\
DELPHI    
  &  $37.36\pm1.00$ &  $19.48\pm0.73$ &
                                  $-1.6\pm1.5$ \\

L3    
  &  $36.3\pm0.9$ &  $18.6\pm0.6$ &
                                  $-1.0\pm0.9$ \\

OPAL  
  &  $39.4\pm1.0$ &  $19.3\pm0.4$ &
                                  $+0.7\pm1.0$ \\
 \hline
\end{tabular}
\end{center}
\end{table}

The mean multiplicities, \nQQQQ, \nQQLV, and \Dnch\ measured by the LEP
collaborations are shown in Table~\ref{tab:nch}. Although the ALEPH mean
multiplicities should not be compared directly with other data, the
associated difference may be more meaningfully compared. The LEP average
multiplicity difference obtained is:
\begin{displaymath}
 \mathrm{LEP}\;\; \Dnch = +0.20\pm0.50 \mathrm{(stat.+syst.)},
\end{displaymath}
This is consistent with there being no change in the \nQQQQ\ due to colour
reconnection effects.  As the systematics considered and the dominant source
varied between the collaborations, this average assumes all systematics were
uncorrelated. If, instead, the smallest overall systematic estimated by any
collaboration is considered as fully correlated, the average value obtained
is $\Dnch = +0.15\pm0.80 \mathrm{(stat.+syst.)}$.  Performing the average
while excluding the ALEPH results leads to the same conclusions.

All of the \SK, \HERWIG\ and \ARIADNE\ models are consistent with
data, although some more extreme models, such as the
instantaneous reconnection scenarios in the \SK\ model, and also the
\ARIADNE\ model \ARMIII~\cite{bib:ARIADNE} in which gluons having
energies greater than \Gw\ are allowed to interact, are disfavoured.

\begin{table}
\begin{center}
\caption{Dispersions of charged particle multiplicities, errors are
sum of statistical and systematic components}\label{tab:disp}
\vspace{0.2cm}
\begin{tabular}{|l|c|c|c|} \hline
  Expt. &  \nQQQQ  &  \nQQLV &  \Dnch \\
\hline
DELPHI    
  &  $8.24\pm0.51$ &  $5.76\pm0.49$ &
                                  $+0.09\pm0.84$ \\
OPAL  
  &  $8.8\pm0.7$ &  $6.1\pm0.5$ &
                                  $+0.2\pm0.6$ \\
 \hline
\end{tabular}
\end{center}
\end{table}
DELPHI and OPAL also measure the dispersions of the charged multiplicity
distributions, \DQQQQ\ and \DQQLV, and the difference
defined as $\DD=\DQQQQ-\sqrt{2}\DQQLV$. Their corrected results,
shown in Table~\ref{tab:disp}, are consistent with Monte Carlo
expectations and there is no indication for differences in
shape based on the first moments of the multiplicity distributions.

\subsection{Ellis-Geiger Model}
The \EG\ model,~\cite{bib:EG} implemented in the Monte Carlo program \VNI,
has been studied by the LEP collaborations. ALEPH and OPAL illustrate this
by quoting some predictions from \VNI.  It is noted that the \EG\ model as
implemented within \VNI\ has not been tuned recently to describe \Zz\ data.
ALEPH compared the predictions of \VNI\ with other \WW\ event generators,
including detector simulation. Within the acceptance of their charged
tracking and event selection criteria, ALEPH find \VNI\ to give a charged
track distribution that has $\nQQQQ\sim2$ units higher than data and an
r.m.s.\ $\sim1.7$ units broader than in data. They also observe no
dependence~\cite{bib:EG} of \nQQQQ\ on the minimum angle between jets
assigned to different W bosons. OPAL find from hadron level studies without
detector simulation that \VNI\ gives very high charged particle
multiplicities, $\nQQQQ>50$, and also does not reproduce the large predicted
shifts in \nQQQQ\ in any of the colour blind, colour singlet or colourful
scenarios.~\cite{bib:EG} These studies were carried out at $\roots=183$~GeV,
but the same conclusions hold at $\roots=166$~GeV, the lowest centre-of-mass
energy allowed by \VNI\ for \WW\ production.  For these reasons, the
collaborations do not use the \EG\ model as currently implemented in \VNI\
to estimate possible systematic effects on the W boson mass.

\subsection{Thrust}
\begin{figure}
\center
\psfig{figure=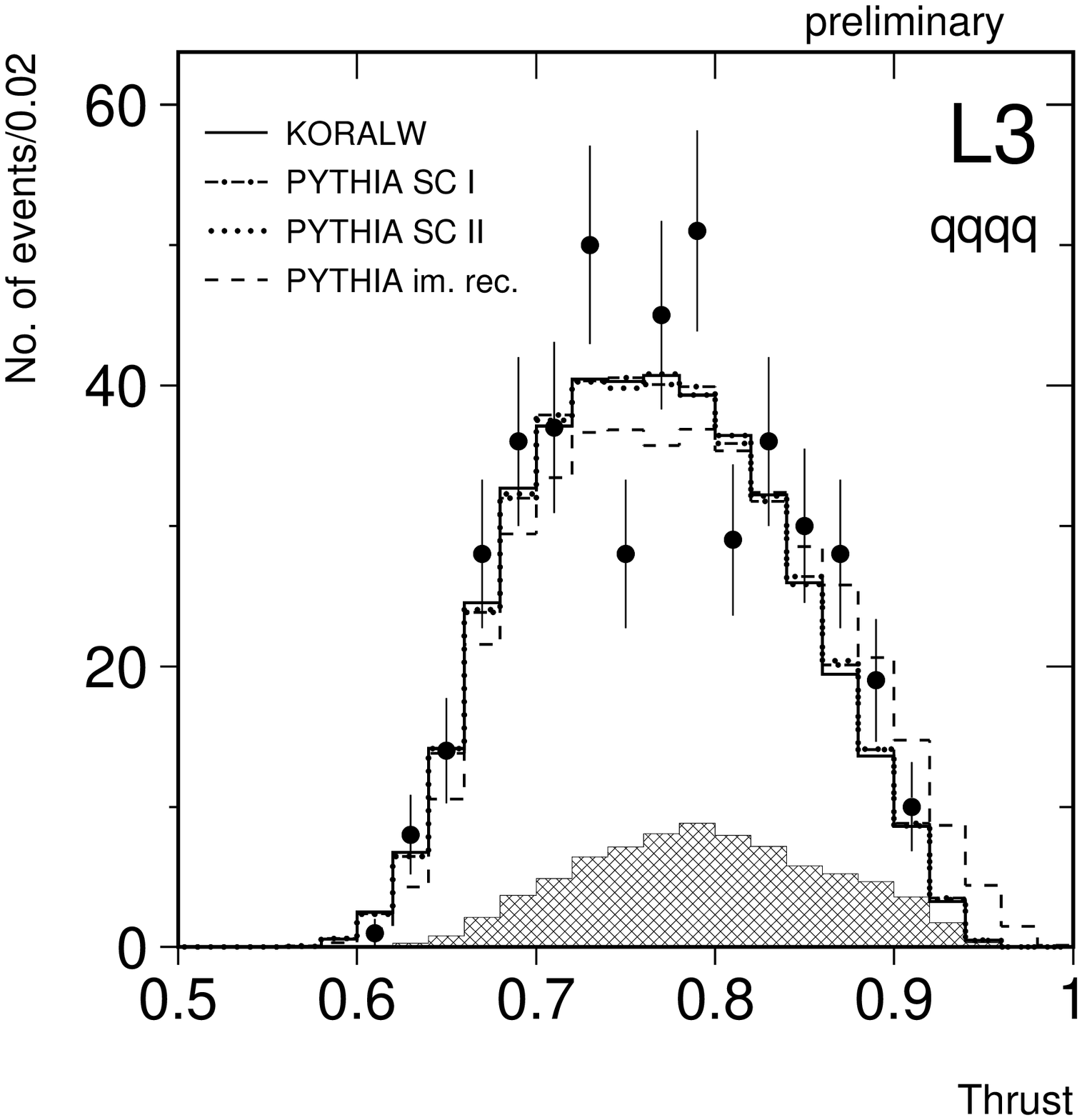,width=8cm}
\caption{Observed thrust, \WWqqqq\ events.}
\label{fig:thr_l3}
\end{figure}
 The thrust, $T$, distribution shown in Figure~\ref{fig:thr_l3} from
L3 illustrates how \WWqqqq\ events are more spherical than the \Zqq\
background, which is dominated by two-jet events. Qualitatively,
colour reconnection effects are expected to be enhanced in \WW\ events
where the hadronic cascades from two W bosons overlap, but this is
precisely the background dominated, two-jet like region generally
excluded by the experimental event selections.
\begin{figure}
\center
\psfig{figure=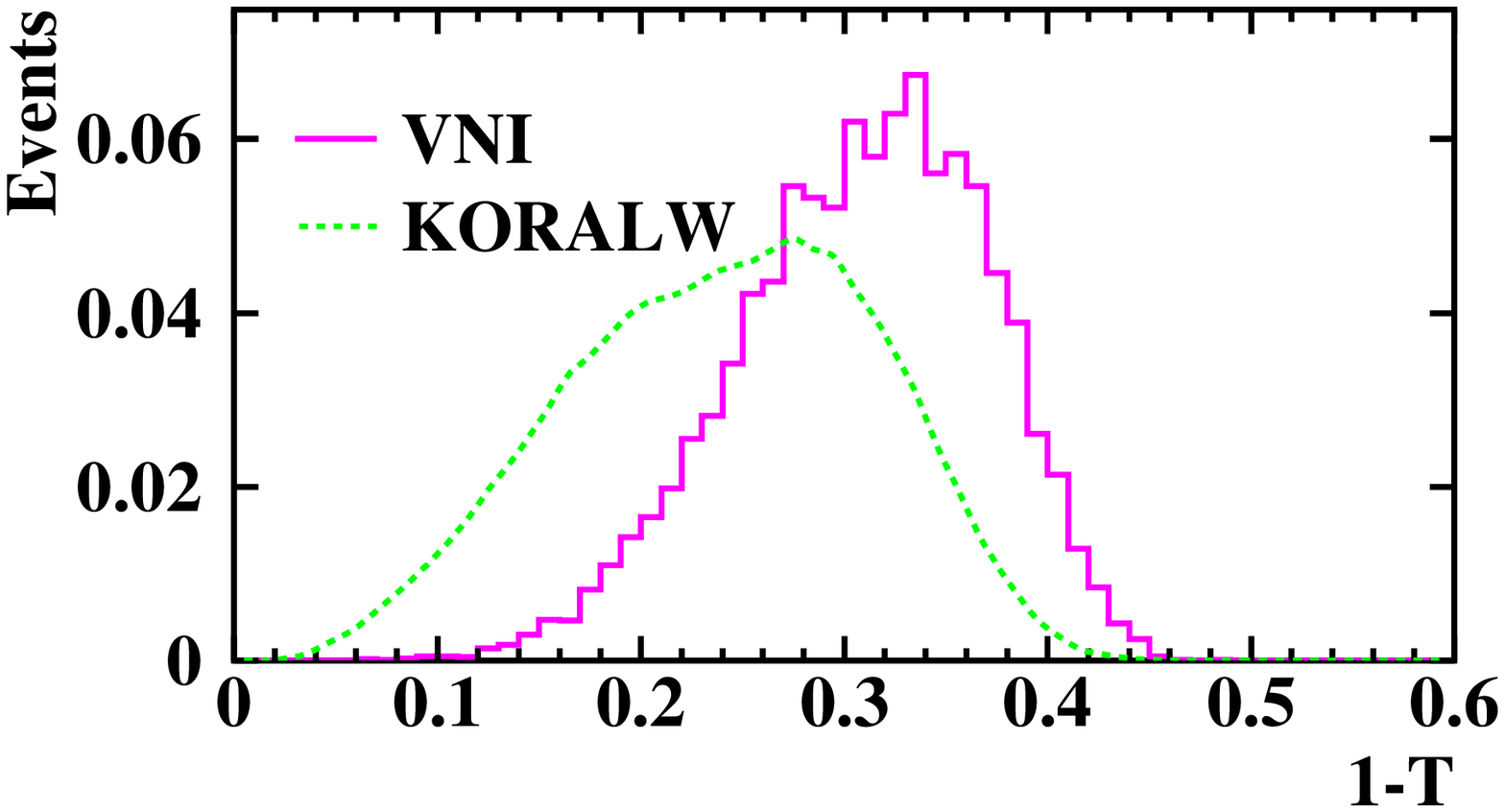,width=8cm}
\caption{Hadron level predictions, \WWqqqq\ events.}
\label{fig:thr_vni}
\end{figure}

It is interesting to compare the hadron level predictions of a
standard \WW\ event generator such as \KORALW\ with the \VNI\
generator in its colour blind scenario. The distribution of $1-T$ is
shown in Figure~\ref{fig:thr_vni} for these two models. It can be seen
that \VNI\ predicts \WW\ events to be more spherical than other
models. While this might influence the efficiency with which such
events are selected, they are topologically more distinct from the
background than predicted by most models, so should fall well within
the acceptance of the detectors.

\subsection{Model Tuning}
\begin{figure}
\center
\psfig{figure=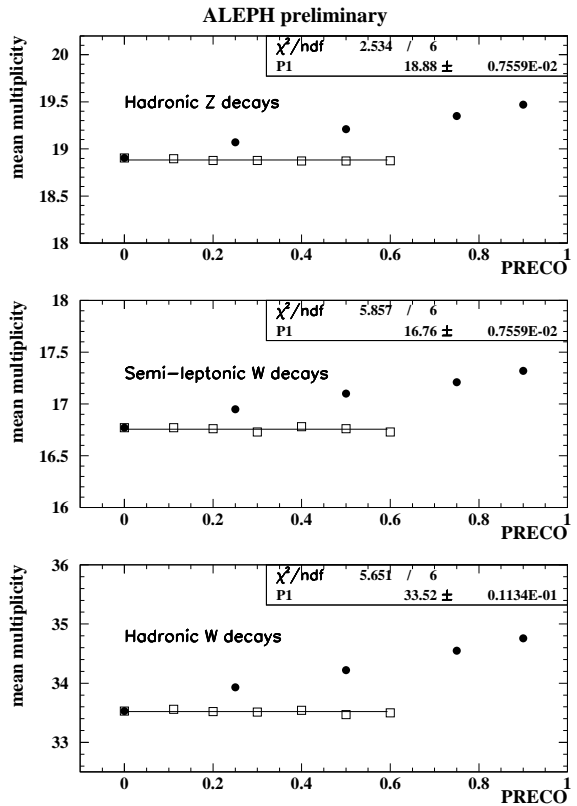,width=8cm}
\caption{Tuning of \HERWIG\ reconnection model.}
\label{fig:hwtune_aleph}
\end{figure}
In general each model must be retuned to data after fixing reconnection
related strength parameters or probabilities. The \SK\ models do not require
such retuning once the \JETSET\ hadronisation model has been tuned. \HERWIG,
\ARIADNE\ and the \EG\ model all require retuning. OPAL use a tuning of
\ARIADNE\ summarised in,~\cite{bib:qgdiffn} while ALEPH describe in detail
their tuning of \HERWIG.  The mean charged multiplicities in \HERWIG\ for
\Zgamma, \WWqqlv\ and \WWqqqq\ events as a function of the reconnection
probability, \texttt{PRECO}, are shown in Figure~\ref{fig:hwtune_aleph} by
ALEPH.  The effect of retuning at each \texttt{PRECO}, and its necessity, is
clear. A good description of the \WW\ events may be achieved after suitable
retuning to \Zz\ data.

\subsection{Alternative Observables}
\begin{figure}
\center
\psfig{figure=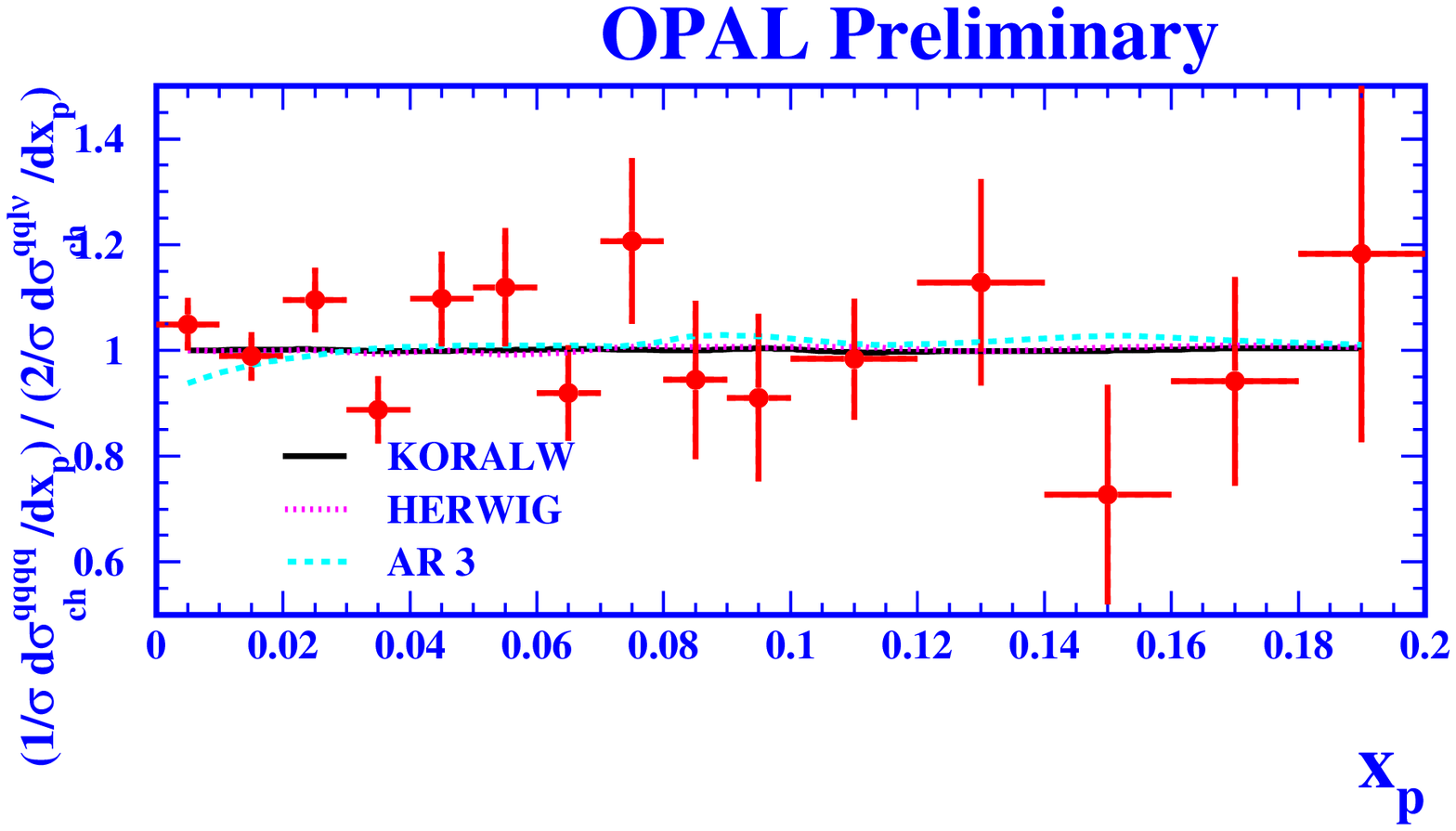,width=8cm}
\caption{Momentum distribution ratio, \WWqqqq\ and \WWqqlv.}
\label{fig:xp_opal}
\end{figure}

\begin{figure}
\center
\psfig{figure=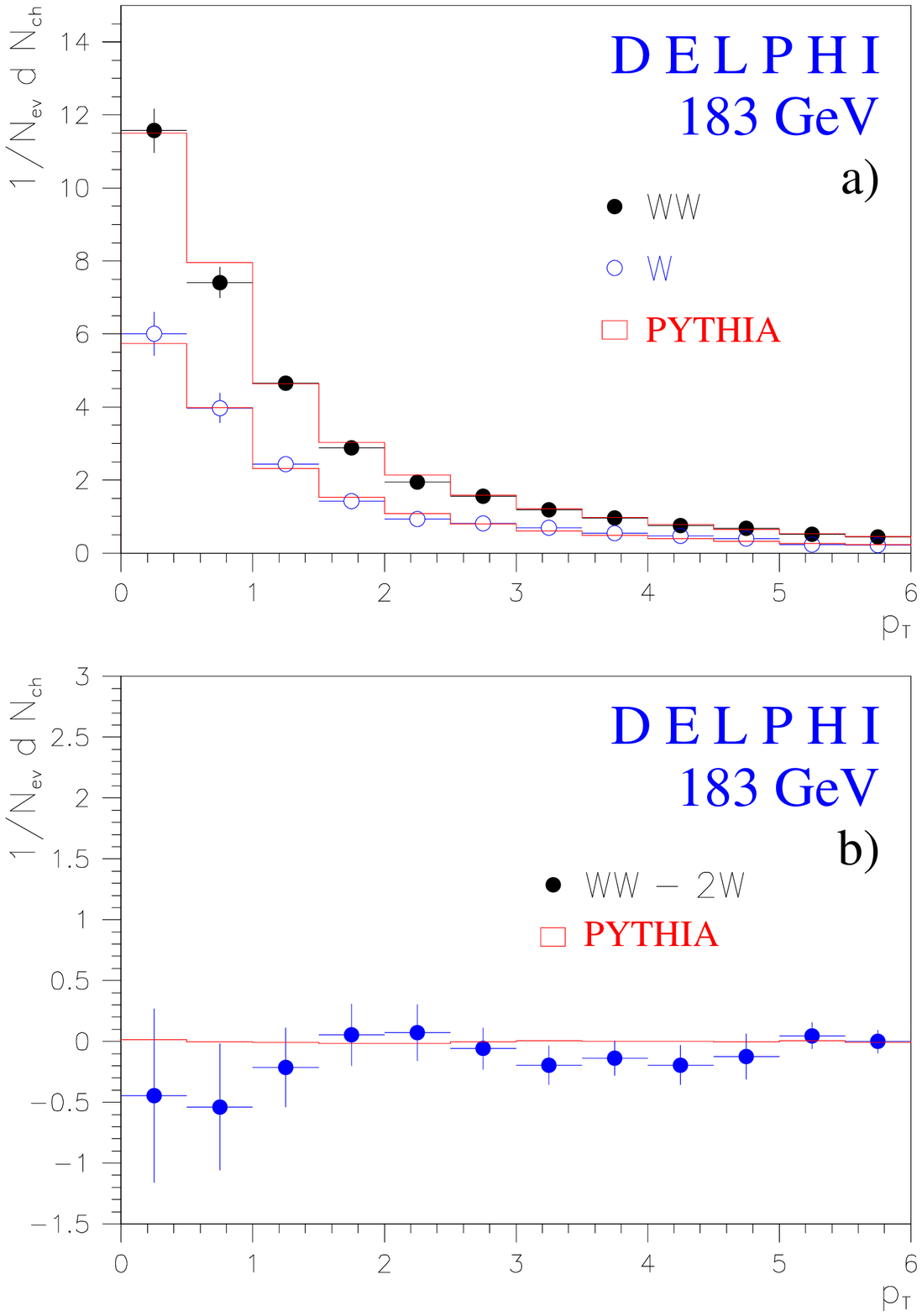,width=8cm}
\caption{Transverse momentum distribution, \WWqqqq\ and \WWqqlv.}
\label{fig:pt_delphi}
\end{figure}

The effects of colour reconnection are generally predicted to be enhanced in
the soft particle region, therefore studies of fragmentation functions are
interesting. Figure~\ref{fig:xp_opal} shows the ratio of the corrected,
scaled momentum distributions for \WWqqqq\ and \WWqqlv\ events, together
with predictions of \WW\ event generators and the colour reconnection model
in which they found the largest effects, \ARMIII. Although this latter
predicts differences which are as large as 5\% in the lowest momentum
interval, it is still consistent with data.  OPAL also measure a variety of
mean event shape variables, all of which are consistent with model
predictions (neglecting \VNI):

\begin{eqnarray*}
 \xpQQQQ  & = & (3.16  \pm 0.05  \pm 0.04)\times 10^{-2} \\
 \xpQQLV  & = & (3.25  \pm 0.07  \pm 0.04)\times 10^{-2} \\
 \Dxp     & = & (-0.09 \pm 0.09  \pm 0.06)\times 10^{-2} \\
 \thrQQQQ & = & 0.240 \pm 0.015 \pm 0.009 \\
 \yQQQQ   & = & 1.017 \pm 0.016 \pm 0.016
\end{eqnarray*}

\noindent
DELPHI show in Figure~\ref{fig:pt_delphi} the transverse momentum
distributions relative to the thrust axes for \WWqqqq\ and \WWqqlv\ events,
and their ratio.  In the semi-leptonic events the thrust axis is calculated
taking the momentum imbalance in each event to be the neutrino
momentum.There is no significant difference found between the data in the
two channels.

\section{Heavy Hadrons}
\begin{figure}
\center
\psfig{figure=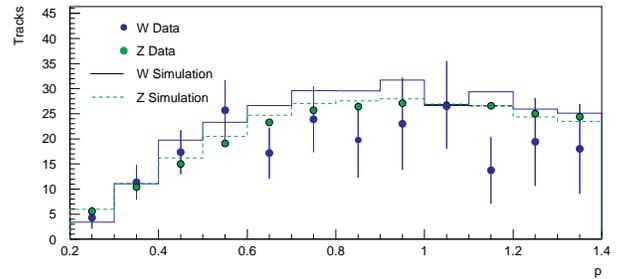,width=8cm}
\caption{Comparison of \WW\ and \Zz\ pair data.}
\label{fig:heavy_delphi}
\end{figure}
It was recently predicted~\cite{bib:SK98} that the effects of colour
reconnection may be further enhanced by restricting analyses to
heavier hadrons such as charged kaons and protons in the low momentum
region, $0.2<p<1.2$~GeV. Numerically this increases the relative size
of the effect, but taking into account the branching fractions to such
species as well as experimental effects such as finite particle
detection efficiency, background and modelling, it is not clear this
will lead to an improved significance in the analyses.  DELPHI use
their TPC, RICH and vertex detector for particle identification,
attaining tagging efficiencies for kaons and protons of order
90--50\%, with purities in the range 90--75\%.  They
observe 181 heavy hadrons in \WWqqqq\ events, 88 in \WWqqlv. After
unfolding for tagging effects this gives:

\begin{eqnarray*}
(\qq\qq)/(2\qq\lnu) & =& 1.03\pm0.15 \mathrm{(stat.)}.
\end{eqnarray*}

\noindent
The statistical uncertainty on this measurement seems unlikely to be less
than 5\% even with the nominal LEP2 integrated luminosity of 500~pb$^{-1}$.
To reduce the statistical uncertainty on these studies, they also use \Zz\
calibration data as a reference sample. Pairs of \Zz\ events, after
boosting, are used to emulate \WWqqqq\ events, and a direct comparison is
made with genuine \WWqqqq\ data. The two data samples and the corresponding
Monte Carlo predictions are shown in Figure~\ref{fig:heavy_delphi}.  The
qualitative agreement is reasonable, although there may be differences
between \WW\ data and predictions for $p>0.6$~GeV.  The final result of
their analysis is the ratio:
\begin{eqnarray*}
\langle n^{\mathrm{heavy}} \rangle, \;\;
\frac{\WW}{\Zz\Zz}  &  =  &  0.870\pm0.090 
     \mathrm{(stat.)}\pm0.044 \mathrm{(syst.)}.
\end{eqnarray*}

\section{W Mass Biases}
Updated bias estimates for the effects of colour reconnection
on the W mass measurement were presented by ALEPH, DELPHI and OPAL,
as summarised in Table~\ref{tab:random}. It is noted that ALEPH and OPAL
estimates are using their respective default \Mw\ analyses, while 
the DELPHI results are a variant on their main \Mw\ analysis.
The bias estimates from \HERWIG\ and \ARIADNE\ are made after
these models have been retuned to describe \Zz\ data at $\roots=91$~GeV.
As a cautionary note, there are non-negligible statistical 
uncertainties on all of these bias estimates which must be reduced
for future analyses. No bias estimates are presented for the \EG\
model for reasons discussed earlier.
\begin{table}
\begin{center}
\caption{Model dependent, colour reconnection bias estimates
 for \Mw\ determination in \WWqqqq\ channel (statistical uncertainties).
 $\dag$: fast detector simulation, $\ddag$: modified version of model.}
 \label{tab:random}
\vspace{0.2cm}
\begin{tabular}{|l|c|c|c|} \hline
  Expt. &  ALEPH  &  DELPHI$^{\dag}$
                          &  OPAL \\
\hline
\SKI 
  &  $+25\pm21^{\ddag}$        &  $+40\pm12$  &
                                  $+50\pm17$ \\
\SKII 
  &  $+5\pm15$ &  $-$          &
                                  $+17\pm17$ \\
\SKIIpr 
  &  $+17\pm15$ &  $+3\pm10$   &
                                  $+18\pm17$ \\
GH 
  &  $-$        &  $+13\pm18$   &
                                      $-$     \\

HW 
  &  $-31\pm25$ &  $-$         &        \\

\ARMII
  &  $-$        &  $-$         & $+73\pm18$ \\

\ARMIII
  & $-$         & $-$        &
                                          $+146\pm18$ \\
 \hline
\end{tabular}
\end{center}
\end{table}

\section{Summary}
Studies of colour reconnection are maturing, in particular the area of
Monte Carlo tuning is being addressed. There is no significant
difference observed in the charged particle multiplicity of a single W
boson produced in either a \WWqqqq\ or a \WWqqlv\ event.  Most models
are consistent with data based on an integrated luminosity of $\sim
55$~pb$^{-1}$ per collaboration. Identified particle studies in
progress pose an interesting experimental challenge. The \EG\ model,
as currently implemented and tuned, is not used for \Mw\ bias
estimates as it does not describe the data.  Finally, the 1998 data
should allow some models to be excluded leading to better control of
the colour reconnection systematic on \Mw\ measurements.

\section*{Acknowledgements}
Work supported by PPARC grant GR/L04207.  I would like to thank
colleagues in the LEP collaborations, particularly R.~Jones,
B.~Pietrzyk., P.~de~Jong, J.~Mnich, C.~Matteuzzi, N.~Neufeld and
M.F.~Watson.  It is with deepest sorrow that I note the recent, tragic
death of Klaus Geiger, who will be missed by many.

\end{document}